\documentclass[twocolumn,showpacs,preprintnumbers,amsmath,amssymb,prl]{revtex4}

\usepackage{graphicx}
\usepackage{dcolumn}
\usepackage{bm}

\begin{document}

\title{Possible high-temperature superconductivity in multilayer graphane: can the cuprates be beaten?}
\author{V.M.~Loktev$^{a}$, V.~Turkowski$^{b}$\altaffiliation{Corresponding author, e-mail address: vturkows@mail.ucf.edu}}

\affiliation{$^{a}$ Bogolyubov Institute for Theoretical Physics, National Academy of Sciences of Ukraine, Metrologichna str. 14-b, Kyiv, 03680 Ukraine\\
$^{b}$ Department of Physics and NanoScience and Technology Center, University of Central Florida, Orlando, FL 32816}

\date{\today}

\begin{abstract}
We analyze a possible superconductivity in the hole-doped system of layered hydrogenized graphene by taking into account thermal fluctuations of the order parameter. 
In particular, we demonstrate that in the one-layer case the values 
of the high mean-field (MF) critical temperature $T_{c}^{MF}\sim 80-90K$, predicted recently by Savini et al \cite{Savini}, 
do not alter significantly due to the fluctuations, and the Berezinskii-Kosterlitz-Thouless (BKT) critical temperature
of the vortex superconductivity is almost the same as the MF temperature at doping 0.01-0.1. 
We show that in the case of multilayer system, when the coupling between the layers stabilizes the superconducting phase
in the form of fluxon superconductivity, the critical temperature $T_{c}$ can increase dramatically
to the values $\sim 150K$, higher than the corresponding values in cuprates under ambient pressure.  
\end{abstract}

\pacs{74.20.-z, 74.78.-w, 74.40.+k, 73.63.-b}

\maketitle

{\it Introduction.}-- Graphene and its modifications are considered to be 
among the most promising elements of the future technological devices,
in particular due to their unusual electronic transport properties (see, e.g.,
a review Ref.~\cite{CastroNeto}). Possible superconductivity (SC) in graphene
remains one of the most intriguing questions.
While there are some experimental evidences of SC in graphite and graphite-sulfur composites, 
\cite{Kopelevich,daSilva,Moehlecke}
in the two-dimensional (2D) graphene system such a possibility of SC was studied only theoretically 
in frameworks of different models (see, e.g., Refs. 
\cite{Uchoa,Zhao,Marino,BlackSchaffer,Honerkamp,Kopnin,Gonzales,Loktev1}), 
and remains in  principle an open question. 
Since different calculations demonstrate that inter-electron coupling in graphene, including the phonon-mediated coupling,
is not too large the critical temperature is also not expected to be too high. Optimistic estimations
include increase of $T_c$ up to ~10K in the doped graphene due to a van Hove singularity in the electron density of states (DOS).
\cite{Gonzales} 
Moreover, this value may be significantly reduced due to thermal fluctuations of the order parameter (OP).\cite{Loktev1} 
Recently, it was predicted by using first-principle calculations,\cite{Savini} 
that the critical temperature can be much higher in graphane, 
a fully hydrogenized version of graphene . Indeed, as the calculations of the authors of paper \cite{Savini}
show, the electron-phonon interaction in this system is much stronger comparing to graphene, due to a Kohn singularity in the phonon 
spectrum. The estimation of the electron-phonon
coupling constant $\lambda\sim 1.3$ led the authors to the conclusion that the critical temperature can be of order 80K-90K at rather 
small values 
of the hole doping, $0.01-0.1$. 
These estimations were based on the BCS expression for the MF critical temperature. However, in principle in the 2D case the MF 
approximation is valid only in the case of highly doped system, when the Fermi energy is much larger than the typical
phonon (Debye) frequency.\cite{Loktev2,Loktev3} At lower values of doping the MF critical temperature is significantly reduced 
due to thermal fluctuations of the SC OP.\cite{Loktev3} 
In this case the real critical temperature, correspond 
to the BKT temperature temperature. Below this temperature the OP 
(its phases) becomes algebraically ordered, forming the so called vortex SC state (see, e.g., a review \cite{Loktev4}). 
Since, in the one-layer graphane these fluctuations may reduce $T_c$, 
in the case of multilayer graphane, the inter-layer coupling which usually leads to an increase of $T_c$ of the coupled vortex
(fluxon) SC phase (see, e.g., Ref.~\cite{Loktev5} and references therein), the high critical temperature
may be even raised above  $\sim 90K$. In this Letter, we systematically analyze the role of the thermal
fluctuations in the doping dependencies of the critical temperature of the vortex and fluxon SC
in the case of one-  and multilayer systems. 
We show that in the first case
the fluctuations do not suppress $T_c$ significantly at doping lager than $0.01$, and in the second case 
the inter-layer coupling may lead to a significant increase of $T_c$ to the values 
higher than the cuprate critical temperature $135K$, the maximal $T_{c}$ under normal pressure
known so far.

{\it Model and main equations.}--The secondary-quantization effective SC Hamiltonian for the multilayer graphane 
can be written in the following form:
\begin{eqnarray}
H(t)=\sum_{l\sigma}\int d^{2}r\psi_{l\sigma}^{\dagger}(\tau,{\bf r})
\left[ -\frac{{\bf \nabla}^{2}}{2m}+2t-\mu\right]\psi_{l\sigma}(\tau,{\bf r}) \nonumber \\
-\sum_{l,m,\sigma}t_{lm}\int d^{2}r\psi_{l\sigma}^{\dagger}(\tau,{\bf r})\psi_{m\sigma}(\tau,{\bf r})
+g\sum_{l,\sigma}\int d^{2}r\varphi_{l}(\tau,{\bf r})
\nonumber \\
\times\psi_{l\sigma}^{\dagger}(\tau,{\bf r})\psi_{l\sigma}(\tau,{\bf r})
+\sum_{l}H_{ph}[\varphi_{l}(\tau,{\bf r})], \ \ \ \ \ \ \ \ \ \ \ \ \ \ \ \ \ \ \ \
\label{H}
\end{eqnarray}
where the first and the second terms correspond to the in-plane and out-plane components of the electron kinetic energies, 
the third term corresponds to the energy of electron-phonon interaction and the last term is the free phonon Hamiltonian. 
In Eq.~(\ref{H}),  
$\psi_{l\sigma}^{\dagger}(\tau,{\bf r})$ and $\psi_{l\sigma}(\tau,{\bf r})$
are the Heisenberg creation and annihilation operators of electron (hole) with spin $\sigma =\pm 1/2$ 
in the $l$th layer at point {\bf r} and Matsubara (imaginary) time $\tau$; 
m, $\mu$ and $t$ are the in-plane electron (hole) effective mass, chemical potential and inter-layer hopping energy; 
$t_{lm}=t(\delta_{l,m-1}+\delta_{l,m+1})$ 
is the nearest-neighbor inter-layer hopping matrix;  g is the electron-phonon coupling. 
We consider the simplest dispesionless (Einstein) phonon model,
so in this case the free phonon propagator, which enters in the last term in Eq.~(\ref{H}), 
has the following form in the Matsubara frequency representation: $D(i\Omega_n )=-\omega_0^2/[\Omega_n^{2}+\omega_0^2]$, 
where $\omega_0$ and $\Omega_n=2n\pi T$ (T is temperature and n is an integer number) are the Einstein 
phonon and Matsubara frequencies. The first frequency can be considered as a weighted phonon frequency
in the case of phonons with dispersion. The possible values of $\omega_0$ are discussed below.
Since the one-layer graphane is a direct band semiconductor 
(with the gap $\sim 3.5eV$ near the $\Gamma$-point), the spectrum of weakly-doped free carriers can be
approximated by the standard ${\bf k}^2$ expression (contrary to the linear Dirac fermion spectrum in the cases 
pure and weakly-doped graphene). 
The effective masses of the doped heavy and light holes and electrons can be aproximated by $m_{hh}=0.64m_{e}$, $m_{lh}=0.22m_{e}$ 
and $m_{le}=m_{e}$, correspondingly ($m_{e}$ is bare electron mass).\cite{Cudazzo,Tokatly} 
Similar to the electron-doped case, in the case of hole doping we shall use one-band approximation with the  average effective 
mass $m_{h}=2/[1/m_{lh}+1/m_{hh}]=0.32m_{e}$. We choose the value of the coupling g such that the BCS coupling 
$\lambda =mg^2/4\pi$ is equal to 1.3 in the hole-doped case.\cite{Savini} 
In the electron case, the corresponding constant must be much larger, due 
to larger effective electron mass. The value of the phonon frequency $\omega_0$ 
can be estimated as $\sim 0.015eV$ in order to reproduce the average BCS value 
for the MF critical temperature $T_c=1.14\omega_0\exp [-1/\lambda]\sim 90K$ obtained in Ref.~\cite{Savini}. 
The value of the remaining free parameter in Hamiltonian (\ref{H}), the interlayer hopping $t$,
can be estimated from the values of the Slater-Koster hopping parameters 
between the ss-, sp- and pp-orbitals $\sim \alpha\hbar^2/m_ed^2$, where d is the inter-atomic 
(inter-layer) distance and $\alpha\sim 1$ is an orbital-dependent coefficient
(see, e.g., Ref.~\cite{Harrison}). 
Using a recent van der Waals-Density Functional Theory  result 
for the distance between the centers of masses of two neighboring layers of graphane $d=4.65\AA$ ,\cite{Rohrer} 
one can assume $t\sim 0.1-0.3eV$. In order to derive 
the equations for the SC OP 
$\Phi_{l}(\tau,{\bf r}_{1},{\bf r}_{2})
=\langle \psi_{l\uparrow}(\tau,{\bf r}_{1})\psi_{l\downarrow}(\tau,{\bf r}_{2})\rangle$, 
$T_{c }$ and the chemical potential as functions of the carrier density and the carrier-phonon coupling, 
one can use the expression for the thermodynamical potential of the system as a functional of  
$\Phi_{l}(\tau,{\bf r}_{1},{\bf r}_{2})$. We consider the case of the s-wave (isotropic) pairing
and weak thermodynamical fluctuations of OP. The last approximation is valid in the case when the temperature
is not too low, which is definitely correct when T is close to $T_{c}$ and the doping is not extremely low. 
Then, one can show that the OP can be approximated by 
$\Phi_{l}({\bf r}_{1},{\bf r}_{2})\simeq \Delta \exp [i\theta_l({\bf r}_{1},{\bf r}_{2})/2)]$, 
where $\Delta$ is the modulus of the OP (superconducting gap), and $\theta_{l} ({\bf r})$ is its phase. 
This phase is proportional to the sum of the phases of the carrier operators
$\psi_{l\uparrow}(\tau,{\bf r}_{1})$ and $\psi_{l\downarrow}(\tau,{\bf r}_{2})$ 
and depends on the center-of- mass coordinate of the Cooper pair. The thermal phase fluctuations are dominant
comparing to the fluctuations of $\Delta$, therefore we shall consider only fluctuations of $\theta$
assuming that $\Delta$ is constant (see, e.g., Ref.~\cite{Loktev4}). We also neglect the temporal dependence
of the OP (which is important in the quantum fluctuation regime at $T\rightarrow 0$).
Then, the thermodynamic potential has the following form in the second order (hydrodynamic) approximation
in the fluctuations of the order parameter: 
$\Omega (\mu ,\Delta ,\nabla\theta ,T)=\Omega^{pot} (\mu ,\Delta ,T)
+J(\mu ,\Delta ,T)/2\sum_{l}\int d^{2}r (\nabla\theta_{l}({\bf r}) )^{2}
+J_{||}(\mu ,\Delta ,T)/2\sum_{l}\int d^{2}r [1-\cos (\theta_{l}({\bf r})-\theta_{l-1}({\bf r}) )^{2}$ ,
where $\Omega^{pot}$ is the expressions for the MF (or BCS) part of the thermodynamic potential,
and $J(\mu ,\Delta ,T)$ and $J_{||}(\mu ,\Delta ,T)$ are the in-plane and inter-plane stiffnesses of the SC phases.
The explicit expressions for these parameters can be found, for example, in Ref.~\cite{Loktev5}.  
In the MF approximation, variation of this potential with respect to $\Delta$ and $\mu$ (neglecting the SC phases) 
leads to the BCS-like equation for the SC gap and the number of particles equation:
\begin{eqnarray}
1\simeq \lambda \int \frac{d^{2}kdk_{||}}{2E({\bf k},k_{||})}
\tanh \left(\frac{E({\bf k},k_{||})}{2T}\right) 
\nonumber \\
\times\theta \left( \omega_{0}-|{\bf k}^{2}/2m-\mu | \right) , 
\label{DeltaBCS}
\end{eqnarray}
\begin{eqnarray}
n_f=\int d^{2}kdk_{||}n_{f}({\bf k},k_{||}) ,
\label{nBCS}
\end{eqnarray}
where $E({\bf k},k_{||})=\sqrt{\xi^{2} ({\bf k},k_{||})+\Delta^{2}}$ is the quasi-particle energy in the SC state
and  $\xi ({\bf k},k_{||})={\bf k}^{2}/2m+2t-2t\cos (dk_{||})-\mu$  is the free hole (electron) spectrum.
$$
n_{f}({\bf k},k_{||})=1-\frac{\xi ({\bf k},k_{||})}{E({\bf k},k_{||})}\tanh \left(\frac{E({\bf k},k_{||})}{2T}\right)
$$
is the charge density distribution in the momentum space.
Solution of the system of Eqs.~(\ref{DeltaBCS}) and (\ref{nBCS}) at $\Delta =0$ gives one the dependence of the MF critical temperature
$T_{c}^{MF}$ on the model parameters. This solution as well as the solution for the real critical temperature $T_{c}$
which is defined by the stiffness parameters $J$ and $J_{||}$ in the case of one- and multi-layer systems will be discussed
below.

{\it One-layer case.}--The solution of the system of MF equations (\ref{DeltaBCS}), (\ref{nBCS}) 
leads to the following approximate dependence of the SC temperature on $n_{f}$:
$$
T_c^{MF}\ln T_c^{MF}/\epsilon_F =\omega_0 e^{-2/\lambda} , \epsilon_F<\omega_0 
$$
$$
T_c^{MF}=1.14\omega_0 e^{-1/\lambda}, \epsilon_F> \omega_0 , 
$$
which becomes exact at $\epsilon_F<<\omega_0$ and $\epsilon_F<\omega_0$, correspondingly.
Since in the 2D case the Fermi energy is connected with the particle density 
by the following relation $\epsilon_F=\hbar^{2}\pi n_f/m$, one can estimate 
that the critical value of the carrier density above which the 
MF temperature reaches the doping-independent BCS value  $1.14\omega_0 e^{-1/\lambda}$
from $\epsilon_F=\omega_0\sim 0.015eV$. This gives
the critical number of doped electrons per atom:  $N_{atom}=0.0017m_h/m_e$ 
(we used the value $v_{cell}=5.4\AA$ for the volume of the unit cell in $n_f=2N_{atom}/v_{cell}$).
Since the effective hole mass 
is smaller than the bare electron mass, one can conclude that the BCS regime in the one-layer case takes place at doping much 
less than 0.01, and that estimations of the MF temperature obtained for doping from 0.01 to 0.1 
in Ref.~\cite{Savini} are correct. As it was mentioned above, there is no long-range SC order in 2D case, 
which means that the OP cannot be constant except the T=0 case. Similar to the XY spin model, there possible 
an algebraic order in the system 
at some temperature $T_c^{2D}$ below which 
the phases of the superconducting order parameter become algebraically ordered . This temperature 
is defined from $T_c^{2D}=\pi /2 J$. Using the expression for the energy J, one can show that
this temperature is approximately equal to  $\epsilon_F/8$
at low doping and approaches the BCS value $T_{c}^{MF}$ with doping increasing at $\epsilon_{F}>\omega_{0}$,
i.e. the fluctuations are not important at doping larger than $0.01$, and  $T_{c}^{2D}\simeq T_{c}^{MF}\sim 90K$ in this case.
This suggests that the pseudogap phase (PG) i.e. the finite temperature interval between $T_c^{2D}$ and $T_c^{MF}$ 
can be observed only at extremely low doping (see Fig.1a). It is interesting to compare how a much large PG region can merge
above $T_{c}^{2D}=\epsilon_{F}/8=\hbar^{2}\pi n_f/8m$ in some of the cuprates. 
In this case, it was estimated that the effective hole mass can be doping-dependent: below  $N_{atom}=0.1$  
it is of order $15m_e$, and after this value of doping it suddenly drops to $5m_e$ and smoothly decreases with doping increasing.\cite{Kristoffel} 
As it follows from the last equation for $T_{c}^{2D}$, the critical temperature for cuprates 
grows linearly with doping and reaches 86K 
at doping 0.1, i.e. there is a large finitite temperature interval between $T_{c}$ and $T_{c}^{MF}$ (which may be associated with
maximal $T_{c}^{2D}$ at optimal doping) for doping values below a rather large value 0.1.
The main reason for this difference in the PG phases is much larger value of the hole effective mass in cuprates,
which leads to a slower growth of $T_{c}^{2D}$ with doping, 
and therefore $T_{c}^{2D}$ ``meets'' $T_{c}^{MF}$ at much larger values of doping
(we do not discuss here the reason why superconductivity in cuprates starts at finite doping $\sim 0.05$, which is a topic 
of a separate extended discussion).  
\begin{figure}[t]
\includegraphics[width=7.5cm]{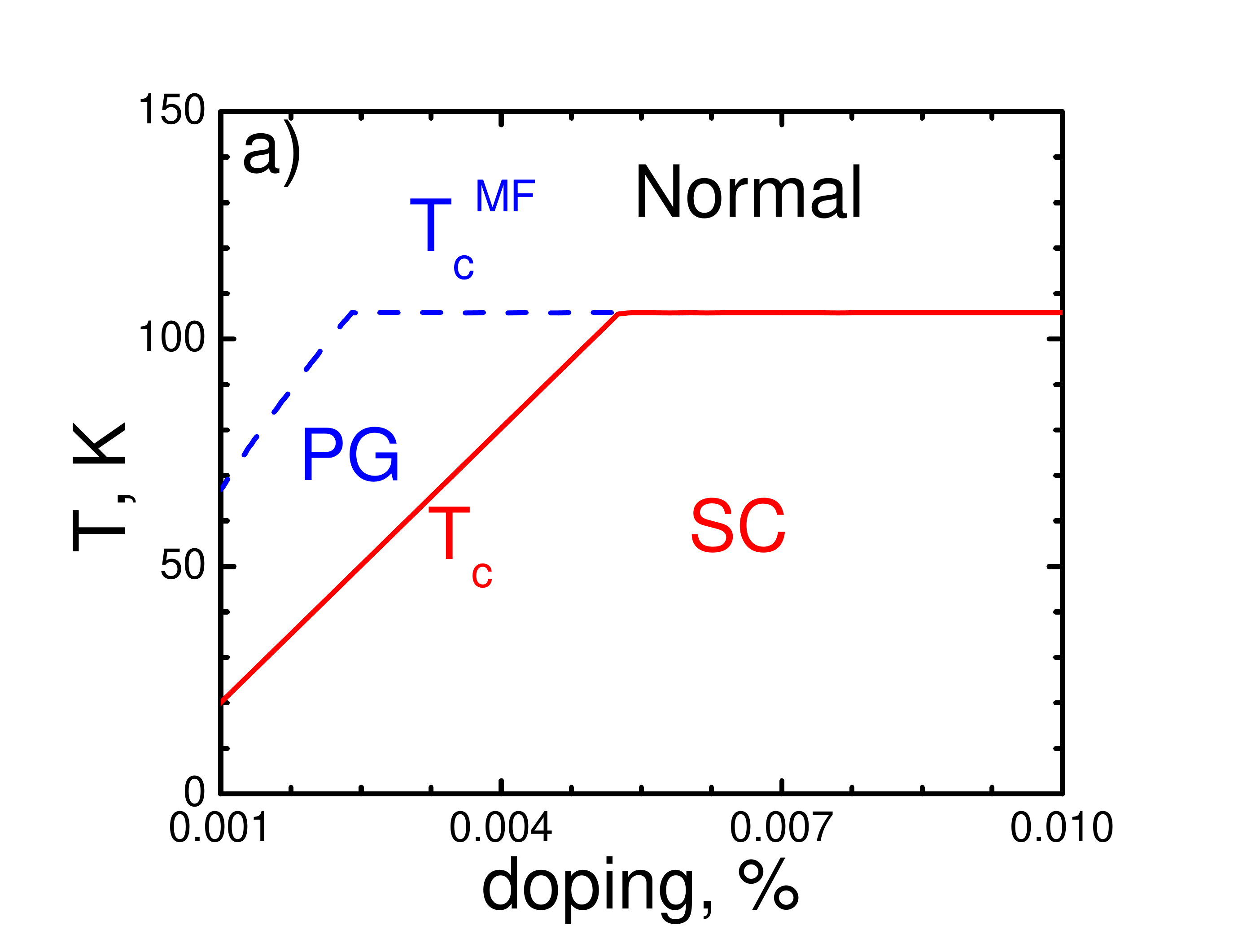}
\includegraphics[width=7.5cm]{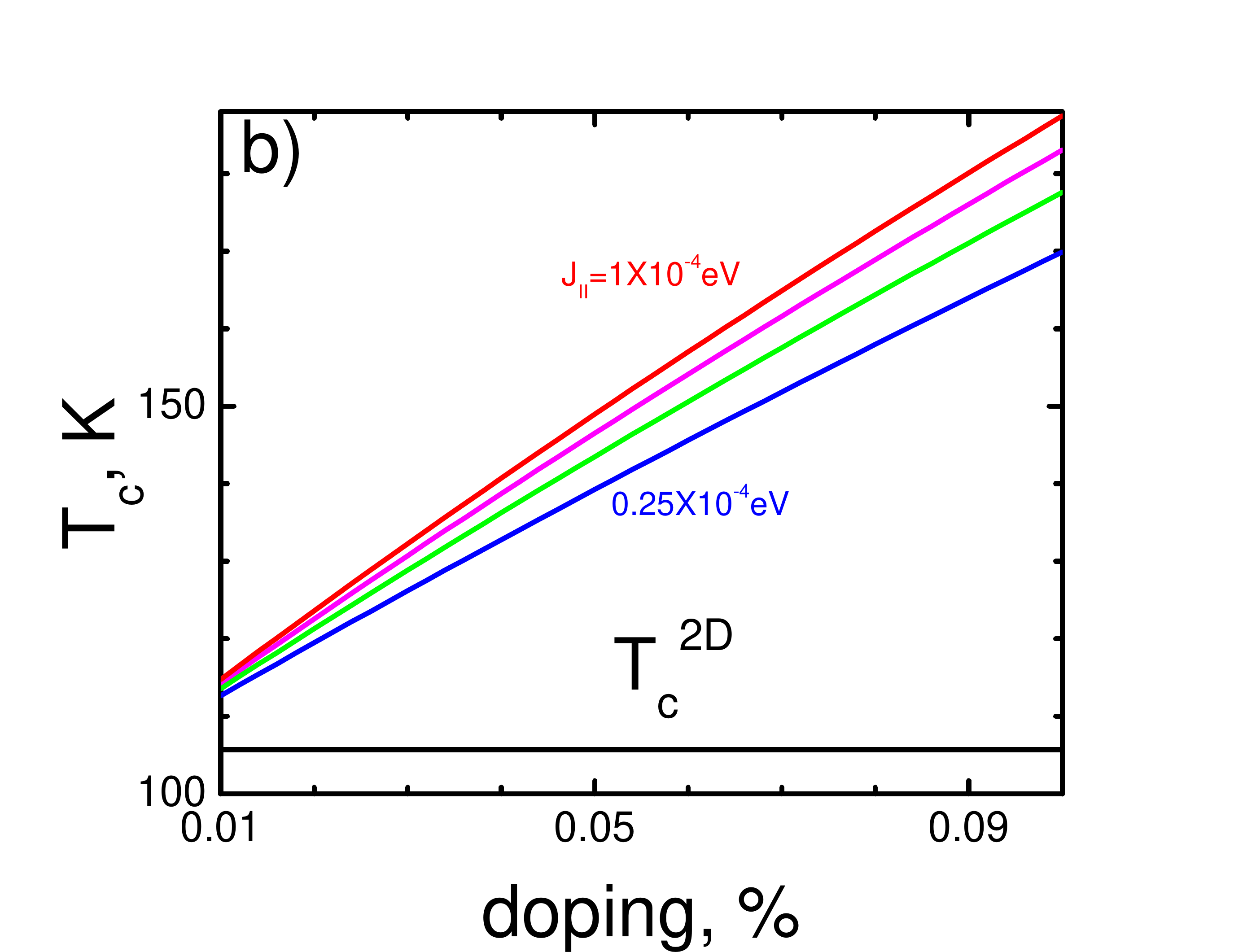}
\caption{\label{fig1} The doping dependence of the critical temperaturein the case of one-layer (a) 
of the multilayered system at different values of the inter-layer stifness $J_{II}$  (b). The model parameters 
used in the calculations are: $\lambda =1.3$, $\omega_{0}=0.015eV$, $m_{h}=0.32m_{e}$.Here and in the next Figure, we have used 
an approximate parabolic hole bandwidth $W=2.7eV$, which corresponds to the maximal doping 0.2.}
\end{figure}
The solution of the system of equations (\ref{DeltaBCS}), (\ref{nBCS}) for $T_{c}^{MF}$ together with approximate solution
$T_{c}^{2D}\simeq\epsilon_{F}/8$ is presented in Fig.1a. As it follows from this Figure,
the MF approximation is valid at values of doping beginning from less than 0.01, which means that the critical temperature 
$T_{c}^{2D}\sim 90K$ can be reached already at doping less than $\sim 0.01$! It is important to notice thatat large doping the solution of the 
gap equation (\ref{DeltaBCS}) with the number of particle constraint  (\ref{nBCS}) leads to a higher value of $T_{c}^{MF}\simeq 105.7K$,
comparing to the BCS solution  $T_{c}^{MF}\simeq 1.14\omega_0 e^{-1/\lambda}\simeq 90K$. The reason for this that the BCS result,
which follows from the equation (\ref{DeltaBCS}) at large doping (when  $\mu\simeq \epsilon_{F}$), is valid only at small values of $\lambda$.

Another interesting result which follows from the MF solution at $T=0$ is the possibility 
of the crossover from the BSC superconductivity-Bose-Einstein condensation (BEC), or superfluidity regime in the system.
Indeed, the chemical potential $\mu\simeq \epsilon_{F}-|\epsilon_b|/2$ becomes negative at doping when the Fermi energy
is lower than half of the pair binding energy $\epsilon_b=-2\omega_{0}\exp (-2/\lambda )$, 
which means that the system is in the BEC regime in this case \cite{Randeria} (see also Ref.~\cite{Loktev4}).
Though, the estimations from $\epsilon_{F}=|\epsilon_b|/2$ with the parameters used above
show that such a crossover can take place at an extremely low doping, less than $0.001$. 

{\it Multi-layer case.}-- It is possible to show, that in the layered system $T_{c}$ can be much larger
than the 2D temperature $T_{c}^{2D}$, and can reach values $8T_{c}^{2D}$ 
at large inter-particle coupling or doping it can be equal $8T_{c}$.
Indeed, as it was shown by Horovitz \cite{Horovitz}, in a rather general case with the inter-layer stiffness  
$J_{||}$, the equation for the critical temperature has the following form:
\begin{eqnarray}
T_c=8T_c^{2D}\frac{E_{c}+T_c^{2D}\ln (T_c^{2D}/J_{||})}{E_{c}+8T_c^{2D}\ln (T_c^{2D}/J_{||})},   
\label{Tcml}
\end{eqnarray}                                                       
where $E_c$ is the loss of the SC condensation energy in the volume $\xi_0^2d$ , 
where $\xi_0=\hbar v_F/\Delta$ is the in-plane coherence 
length (see also Ref.~\cite{Loktev5}, where the density-dependent solution for the cuprates
was analized, and references therein). This energy can be estimated as SC condensation energy density
 $1/2N(\epsilon_F )\Delta^2$   multiplied 
by the volume $\xi_0^2d$.\cite{Taylor} Since the DOS at the Fermi level is equal $N(\epsilon_F)=m_h/(2\pi\hbar^2)$,
one can find $E_c=(\epsilon_F /2\pi )(1\AA^{2}/v_{cell})\simeq \epsilon_F /10.8\pi$ (as above, we choose $v_{cell}=5.4\AA^{2}$). 
Substituting this expression  into Eq.(\ref{Tcml}), and using the fact  
that $T_c^{2D}\sim 90K$ and  $J_{||}\sim (a/d)^{2}J= (a/d)^{2}(2/\pi) T_{c}^{2D}\sim 0.0001eV$, one can show that at doping $\sim 0.01-0.1$ 
the $E_{c}$ terms give a significant contribution in Eq.~(\ref{Tcml}), which may lead to a very large increase
of the critical temperature, $T_{c}\sim 150K$, comparing to the 2D case (Fig.1b).  
Since the result for the $T_{c}$is rather sensitive to the values of the parameters $E_{c}$ and $J_{II}$ , we have studied such a dependence
by varying their values one order of magnitude below and above with respect to  the estimated values $E_{c0}=0.04eV$ (at doping 0.1) and $J_{||0}=0.0001eV$.
The results are presented in Fig.2. As it follows from this Figure, even a rather modest estimation of the values of the parameters results in a significant, by $20\%-50\%$,
increase of the critical temperature comparing to the 2D case.

\begin{figure}[t]
\includegraphics[width=7.5cm]{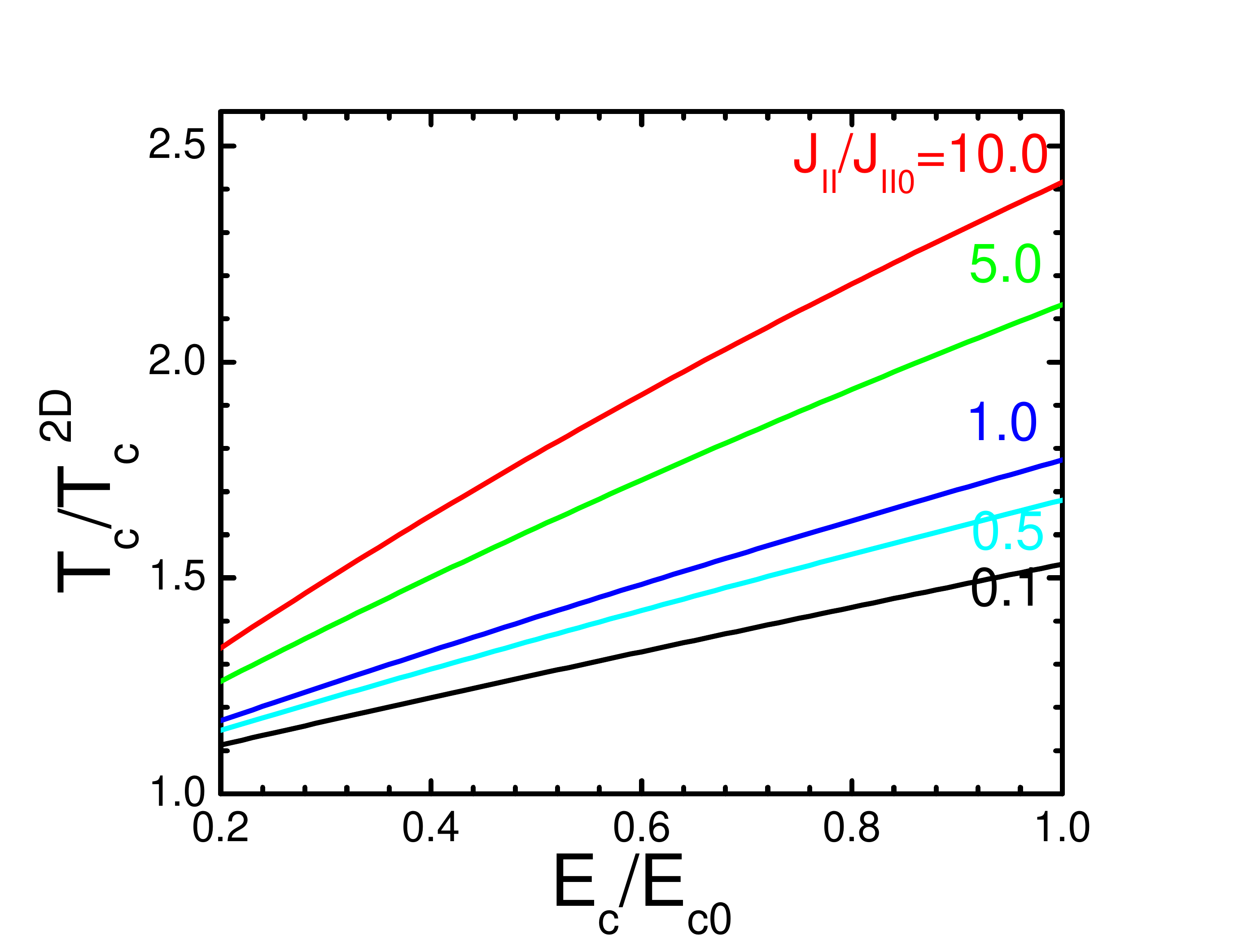}
\caption{\label{fig1} The condensation energy dependence of the critical temperature of the multilayered system (in units of $T_{c}^{2D}=105.7K$)
in the case of different values of the inter-layer stiffness. The reference condensation energy and stiffness are $E_{c0}$ and $J_{II0}$
are given in the text. The other parameters are given in Fig.1}
\end{figure}

{\it Conclusions.}-- In this paper, we have considered superconducting properties of multilayer
graphane by taking into account fluctuations of the order parameter. We have shown that in the single-layer case
the BKT critical temperature which corresponds to the vortex SC is equal to the MF temperature $\sim 100K$
beginning from a rather low values of doping less than $\sim 0.01$. Thus, the high critical temperature due to Kohn
anomaly obtained in the MF approximation for this range of doping 
\cite{Savini} is not altered significantly due to thermal fluctuations. In the case of multilayer graphane we have shown
that the inter-layer coupling, which results in the fluxon SC, may lead to a significant increase of $T_{c}$,
comparing to the single-layer case. Namely, we estimate that the critical temperature may reach values $\sim 150K$,
which is significantly higher than the maximal temperature under ambientl pressure in cuprates.

V.M.L. acknowledges a support through the Special program of the NAS of Ukraine.
V.T. acknowledges a partial support from the Department of Energy under grant number DOE-DE-FG02-07ER15842.

\end{document}